\documentclass[12pt]{iopart}
\usepackage{iopams}

\usepackage[dvips]{graphicx}
\newcommand{\text}[1]{\textrm{\scriptsize #1}}
\newcommand{\mycomment}[1]{}

\begin{document}
\title{Spin polarization of electrons with Rashba double-refraction}
\author{V. Marigliano Ramaglia, D. Bercioux\footnote[1]{Present
address: Institut f{\"u}r Theoretische Physik, Universit{\"a}t
Regensburg, D-93040, Germany}, V.  Cataudella, G. De Filippis and
C.A.Perroni}
\address{{\it Coherentia}-INFM and Dipartimento di Scienze Fisiche
Universit\`a degli Studi {\it Federico II}, Napoli,  I-80126, Italy}
\date{\today}

\begin{abstract}
We demonstrate how the Rashba spin-orbit coupling in semiconductor
heterostructures can produce and control a spin-polarized current
without ferromagnetic leads. Key idea is to use spin-double refraction
of an electronic beam with a nonzero incidence angle. A region where
the spin-orbit coupling is present separates the source and the drain
without spin-orbit coupling. We show how the transmission and the beam
spin-polarization critically depend on the incidence angle. The
transmission halves when the incidence angle is greater than a limit
angle and a significant spin-polarization appears. Increasing the
spin-orbit coupling one can obtain the modulation of the intensity and
of the spin-polarization of the output electronic current when the
input current is unpolarized. Our analysis shows the possibility to
realize a spin-field-effect transistor based on the propagation of
only one mode with the region with spin-orbit coupling. Where the
original Datta and Das device [Appl.Phys.Lett. {\bf 56}, 665
(1990)] use the spin-precession that originates from the
interference between two modes with orthogonal spin.
\end{abstract}

\maketitle

\section{Introduction}

One of the main goal of {\em spintronics} is the production and the
control of spin-polarized currents~\cite{awschalom,zutic}. The
attempts to realize the spin-field-effect transistor (spin-FET)
proposed by Datta and Das~\cite{Datta-Das}, based on Rashba spin-orbit
(SO) coupling \cite{Rashba}, have been unsuccessful because of the
difficulty of spin-polarized injection from a ferromagnetic metal into
semiconductors~\cite{Schmidt}. Despite these obstacles several device
setups based on the Rashba spin-precession have been
proposed~\cite{Kiselev,Koga,Egnes,jiang,frustaglia}. Pala {\it et
al.}~\cite {Pala} have demonstrated the feasibility of Datta and Das
transistor with an all-semiconductor hybrid structure in which the
charge carriers are the holes. A controlled source of spin-polarized
electrons based on a mesoscopic equivalent of an optical polarizing
beam splitter has been proposed~\cite{Ion} . Adiabatic pumping of spin
in presence of a fluctuating electric field, without ferromagnets, has
been discussed~\cite{Govpump}. Finally, Schliemann {\it et
al.}~\cite{schliemann} have studied a spin-FET in which the electrons
are subjected both to Rashba and Dresselhaus SO
interaction~\cite{dresselhaus}. They shows that the balance of the
interaction strengths gives rise to a decoupling of the spin-state and
the momentum direction that allows a non-ballistic spin-FET.

In this paper we present the study of a hybrid system based on Rashba
SO coupling without ferromagnetic contacts. We show that electrons
injected unpolarized from a source are extracted with a partial spin
polarization into a non-ferromagnetic drain. The electron of a
two-dimensional electron gas are injected at an out of normal
incidence angle and the spin-FET operates by means of the spin-double
refraction that appears at the interface between a region without SO
coupling and a region where the SO coupling is
present~\cite{Marigliano}.  The use of spin-double refraction to
produce and control spin-polarized current by means of the appearance
of a limit angle for the refraction has been recently proposed also by
\cite{Khodas} in a system complementary to the our: spin-unpolarized
electrons from a Rashba source traverse a region with a lower SO
coupling and are collected by a drain with a stronger Rashba
coupling. In our case the strength of the SO coupling in the source
and in the drain is zero.  The source and the drain could be realized
by using n$^+$-semiconductors. Our model uses an abrupt interface but
a smooth Rashba field should not change the effects of the spin-double
refraction as the WKB approximation of Ref.~\cite{Khodas} shows.  The
novel feature of our setup is the transmission double step shown in
Fig.~(\ref{figure2}), that is accompanied by the appearance of a spin
polarization. We stress that a modulation of the output current can be
obtained with a spin-unpolarized input current, whereas in the
original Datta and Das~\cite{Datta-Das} proposal the current
oscillation stems out from the difference of phase accumulated along a
path in the Rashba region by the two spin propagating modes. In our
system the current modulation and the spin polarization appear when
{\it only one mode} propagates through the Rashba barrier.

%
%
\begin{figure}
	\centering \includegraphics[width=3.in]{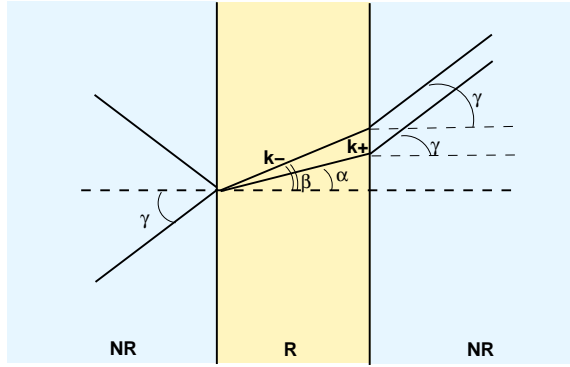}
	\caption{Schematic illustration of the proposed devices. The
	two-dimensional electron gas is divided in three
	region.  In central region a Rashba spin-orbit coupling is
	present. \label{figure1}}
\end{figure}
%
%

We take into account the phase averaging due to the thermal broadening
that tends to wash out the effect of the multiple scattering against
the two interfaces. We show how the resonances that appears when the
electronic beam hits the interface with an angle greater than the
first limit angle, when only one mode traverse the barrier, remain
when the temperature is increased. On the other hand the more rapid
Fabry-Perot oscillation, with rapid changes of the transmission, due
to the propagation of both the modes at low incidence angles are
canceled. The thermal average preserves the halving of the
transmission strictly related to the appearance of the spin
polarization, so that these features do not follow from the multiple
scattering against the two interfaces.

\section{Scattering against a region with SO coupling}

Let us consider a two-dimensional electron gas (2DEG) filling the
plane $(x,z)$. A stripe, where is present the Rashba SO coupling,
divides the plane in three regions as in the Fig.~(\ref{figure1}). In
the inner region R $(0<x<L)$ the Rashba SO coupling term is
%
%
\begin{equation}
	\mathcal{H}_{\text{SO}}=\frac{\hbar k_0}{m}
	\left(\vec{\sigma} \times \vec{p}\right)\cdot \hat{y}  \label{1}
\end{equation}
%
%
where $k_0$ is the SO coupling constant, $\vec{\sigma}$ is the vector
of Pauli matrices, $\vec{p}$ the momentum and $m$ the electron
mass. The strength of the SO coupling can be tuned by external gate
voltages, as it has been experimentally demonstrated
\cite{Nitta,schapers,grundler}. In the outer regions NR ($x<0$ and
$x>L$) there is no SO coupling ($k_0=0$).

Within the R zone there are two spin-split bands $E_{\pm }\left(
k^{\prime }\right) $
%
%
\begin{equation}
E_{\pm }\left( k^{\prime }\right) =\frac{\hbar ^2}{2m}\left( k^{\prime
2}\pm 2k_0k^{\prime }\right) \label{2}
\end{equation}
%
%
where $k^{\prime }=\sqrt{k_x^{\prime 2}+k_z^{\prime 2}}$ is the wave
vector.  The SO interaction can be viewed as due to a magnetic field
parallel to the plane and orthogonal to the wave vector
$\vec{k}^{\prime }$. This magnetic field couples with the
spin-magnetic moment and aligns the spin along the direction
orthogonal to $\vec{k}^{\prime}$~\cite{urRashba}. If $\vec{k}
^{\prime }$ is directed along $x$ then the signs $+$ and $-$ indicate
the ``spin up'' and ``spin down'' states in $z$ direction.

%
%
\begin{figure}
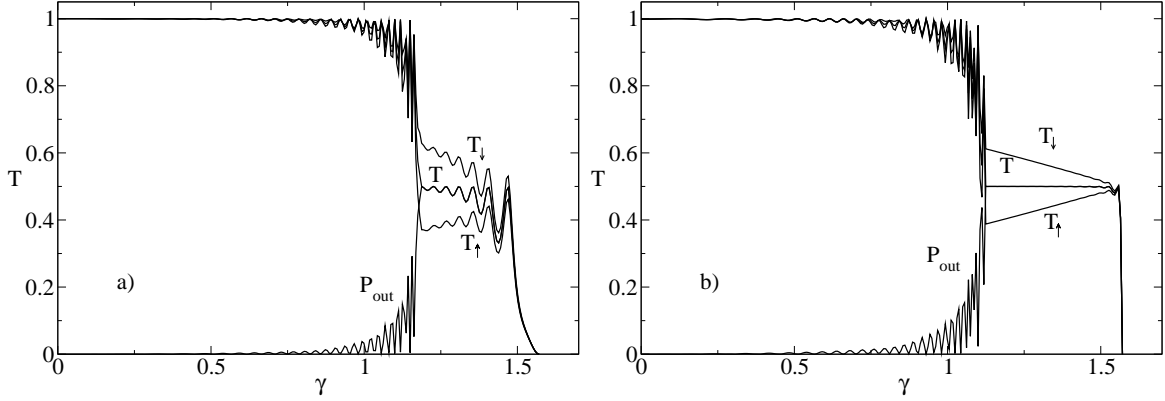

	\centering \includegraphics[width=3.in]{figure2a.eps}
	\includegraphics[width=3.in]{figure2b.eps} \caption{The
	transmission coefficients $T_{\downarrow }\left( \gamma\right)
	,T\left( \gamma \right) ,T_{\uparrow }\left( \gamma \right) $
	as functions of the incidence angle $\gamma $ (in radians) at
	two values of the offset. The length $L$ is 283
	nanometers. The panel a) shows the $1/2$ height resonances,
	where $k_0/k=0.05$ and $u_1/k^2=0.05$. The offset in the panel
	b) is $2k_0/k$ with the largest and the flattest low step as
	obtainable, here $k_0/k=0.05$, $u_1/k^2=0.1$ and $u/k=0$.
	\label{figure2}}

\end{figure}
%
%

The spin split bands may be shifted applying an offset gate voltage
$V_{\text{off}}$ with respect to the source and drain bands $\hbar
^2k^2/2m$. The energy bands (\ref{2}) can be recasted in following form
%
%
\begin{eqnarray}
	E_{\pm }\left( k^{\prime }\right) & = & \frac{\hbar
	^2}{2m} \left( k^{\prime 2}\pm 2k_0 k^{\prime }\right)
	+eV_{\text{off}} = \frac{\hbar ^2}{2m}k^2 \nonumber \\ 
	k^{\prime} & = & \sqrt{k^2-u_1+k_{0}^2} \mp k_0 = 
	k_{\pm } \label{3} \\
	u_1 & = & \frac{2meV_{\text{off}}}{\hbar ^2}.  \nonumber
\end{eqnarray}
%
%

The electron motion within the hybrid NR--R--NR system is supposed as
ballistic and within the Landauer-B\"{u}ttiker regime~\cite{Ferry}.

The single NR-R interface has a transmission coefficient dependent on
the injection angle $\gamma $ and on the incident spin state $\left|
\delta \right\rangle =\cos \delta \left| \uparrow \right\rangle +\sin
\delta \left| \downarrow \right\rangle $ because the Rashba effect
gives rise to the double refraction at the interface with two
orthogonal spin polarizations that simultaneously propagate within the
R zone only when $\gamma \neq 0$~\cite{Marigliano} (out of the normal
incidence). We have already analyzed the case a single NR--R
interface~\cite{Marigliano} and, in this configuration, we have shown
that the sum over all the injection angles reduces, but it does not
cancel, the spin-double refraction effect.

Our aim is to study the conductance of NR--R--NR system in term of the
double interface transmission coefficient $T\left( \delta ,\gamma
\right)$.

In the $x<0$ region we have an incident and a reflected wave 
%
%
\[
{\cos\delta \choose \sin\delta}
e^{ik\left( x\cos \gamma +z\sin \gamma \right) }+ 
{r_{\uparrow} \choose r_{\downarrow}}
 e^{ik\left( -x\cos \gamma +z\sin \gamma \right)}, 
\]
%
%
in the R region ($0<x<L$) we have two propagating and two
counterpropagating waves
%
%
\begin{equation*}
{\cos \alpha /2 \choose -\sin \alpha /2}
t_{+} e^{ik_{+}\left( x\cos \alpha +z\sin \alpha \right) } +
{\sin \beta /2 \choose \cos \beta /2}
 t_{-}e^{ik_{-}\left( x\cos \beta +z\sin \beta \right) }
\end{equation*}
%
%
\begin{equation*}
{\sin \alpha /2 \choose -\cos \alpha /2}
r_{+}e^{ik_{+}\left( -x\cos \alpha +z\sin \alpha \right)}+
{\cos \beta /2 \choose \sin \beta /2}
r_{-}e^{ik_{-}\left( -x\cos \beta +z\sin \beta \right)}
\end{equation*}
%
%
and, finally, the transmitted wave for $x>L$ is 
%
%
\[
{t_{\uparrow} \choose t_{\downarrow}}
e^{ik\left( x\cos \gamma +z\sin \gamma \right)}. 
\]
%
%
The $k_z$ parallel momentum conservation fixes the angular $\alpha $
and $\beta $ directions of $\vec{k}^{\prime }$ 
%
%
\[
\alpha =\arcsin \frac{k\sin \gamma }{k_{+}} \qquad \textrm{and}
\qquad \beta =\arcsin \frac{k\sin \gamma }{k_{-}}. 
\]
%
%
The modes ($+$) and ($-$) have the limit angles $\gamma _0$ and
$\gamma_1$ respectively
%
%
\begin{equation}
\gamma _0=\arcsin \frac{k_{+}}k \qquad \textrm{and} \qquad
\gamma _1=\arcsin \frac{k_{-}}k,  \label{7}
\end{equation}
%
%
that is when $\gamma$ exceeds $\gamma_0$ or $\gamma_1$ the
corresponding mode becomes a decaying wave. The spinors of the wave
function $\psi$ for $0<x<L$ are independent on $\delta$. On the
contrary the spinors of the reflected and the transmitted wave depend
on $\delta$. In the inner region R the four amplitudes
$t_{+},t_{-},r_{+},r_{-}$ vary with $\delta$.  
The single NR-R interface is described by the Hamiltonian 
%
%
\begin{equation}\label{ras6}
\mathcal{H}_{\text{NR-R}}
=\vec{p}\frac{1}{2m(x)}\vec{p}+\frac{k_0(x) m(x)}{\hbar^2} 
\left( \sigma_z p_x -\sigma_x p_z\right) -i\sigma_z \frac{1}{2} \frac{\partial
k_0(x)}{\partial x}+ u \delta(x).  
\end{equation}
%
%
We assume that the mass and the strength of SO coupling are
piecewise constant
%
%
\begin{eqnarray}\label{ras7}
\frac{1}{m}(x) &=& \frac{\vartheta(-x)}{m_\text{NR}} +
\frac{\vartheta(x)}{m_\text{R}}  \\ 
k_0(x) &=& k_0~ \vartheta(x) ,  \nonumber
\end{eqnarray}
%
%
where $\vartheta(x)$ is the step function. To have a model as simple
as possible, we assume that the electron effective mass in the NR
region $m_\text{NR}$ and the electron effective mass in the R
$m_\text{R}$ region are equal. The third term in (\ref{ras6}) is
needed to get an hermitian operator $\mathcal{H}_{\text{NR-R}}$.
The fourth term regulates the transparency of the interface and
describes insulating barriers separating the semiconductors.
The matching conditions at $x=0$ and $L$
%
%
\[
\left\{ 
\begin{array}{r c l}
\psi \left( 0+\right) -\psi \left( 0-\right) & = & 0 \\ 
\psi \left( L+\right) -\psi \left( L-\right) & = & 0 \\ 
\partial _x\psi \left( 0+\right) -\partial _x\psi \left( 0-\right) & = & 
\left( u-ik_0\right) \psi \left( 0\right) \\ 
\partial _x\psi \left( L+\right) -\partial _x\psi \left( L-\right) & = & 
\left( u+ik_0\right) \psi \left( L\right)
\end{array}
\right. 
\]
%
%
provide a linear system for the eight quantities $r_{\uparrow
},r_{\downarrow },t_{\uparrow },t_{\downarrow },$
$t_{+},t_{-},r_{+},r_{-}$. The R region behaves as a resonant cavity
whose action can be reinforced by the couple of additional Dirac-delta
potentials. The strength $u$ of those controls the interfaces
transparency.

To avoid ferromagnetic leads we consider unpolarized electrons
injected into the NR--R--NR system. The unpolarized statistical
mixture at $x=0$
%
%
\[
\rho_\text{in} = 
\frac{1}{2} \left| \uparrow \right\rangle \left\langle \uparrow
\right| +\frac{1}{2} 
\left| \downarrow \right\rangle \left\langle \downarrow \right| 
\]
%
%
becomes the density matrix $\rho _{\text{out}}$ at $x=L$
%
%
\begin{equation}
\rho _{\text{out}}=\frac 12T_{\uparrow }\left| 1\right\rangle \left\langle
1\right| +\frac 12T_{\downarrow }\left| 2\right\rangle \left\langle 2\right|
\label{dario_one}
\end{equation}
%
%
where $T_{\uparrow }=\left| t_{\uparrow \uparrow }\right| ^2+\left|
t_{\downarrow \uparrow }\right| ^2$ is the coefficient for incoming
spin up state and $T_{\downarrow }=\left| t_{\uparrow \downarrow
}\right| ^2+\left| t_{\downarrow \downarrow }\right| ^2$ is that for
the incoming spin down state~\cite{note1}.  The spinors in the
operator (\ref{dario_one}) are
%
%
\begin{equation}\label{dario_ol}
\left| 1\right\rangle =\frac {1}{\sqrt{T_{\uparrow }}}
{t_{\uparrow \uparrow } \choose t_{\downarrow \uparrow}}
\quad\textrm{and}\quad \left| 2 \right\rangle =\frac{1}{\sqrt{
T_{\downarrow }}}
{t_{\uparrow \downarrow} \choose t_{\downarrow \downarrow}}
\end{equation}
%
%
corresponding to input spin up and down respectively. The transmission
coefficient of the unpolarized electrons is 
%
%
\[
T=\left( T_{\uparrow }+T_{\downarrow }\right) /2. 
\]
%
%
We note that $\rho _{\text{out}}$ can be represented in terms of the
output polarization $\vec{P}$ as
%
%
\begin{equation}
\rho _{\text{out}}\left( \gamma \right) =\frac{1}{2}\left(\mathbf{1}+\vec{P}\left(
\gamma \right) \cdot \vec{\sigma}\right)  \label{8}
\end{equation}
%
%
where $\vec{P}$ is the average of $\vec{\sigma}$: 
%
%
\[
\vec{P}=<\vec{\sigma}>=\Tr\left[ \rho _{\text{out}}\vec{
\sigma}\right]. 
\]
%
%
The modulus of $\vec{P}$ gives the degree of polarization in the
output. A simple calculation shows that the modulus of the
polarization $P_{\uparrow }$ of the spinor $\sqrt{T_{\uparrow }}\left|
1\right\rangle $ is $P_{\uparrow }\equiv T_{\uparrow }$, whereas the
modulus of the polarization $P_{\downarrow }$ of $\sqrt{T_{\downarrow
}}\left| 2\right\rangle $ is $P_{\downarrow }\equiv T_{\downarrow
}$. Finally, when the input state is unpolarized the output state is
partially polarized. In particular, for $\gamma >\gamma _0$ we get
%
%
\begin{equation}
\left| \vec{P}_{\text{out}}\right| =\frac{1}{2}\left( T_{\uparrow
}+T_{\downarrow }\right) =T \quad \textrm{for} \quad \gamma >\gamma_0
\label{9}
\end{equation}
%
%
since $\left| \left\langle 1|2\right\rangle \right| $ goes very
quickly to 1. On the contrary when the incidence angle $\gamma $ is
lower than $\gamma _0,$ $\left| \vec{P}_{\text{out}}\right| <T$ and
the polarization vanishes when $\gamma $ goes to zero.

Finally for $\gamma >\gamma _0$ some resonances appear (see
Fig.~\ref{figure2}) for  which 
%
%
\begin{eqnarray*}
R_{\uparrow } &=&T_{\downarrow } \\
R_{\downarrow } &=&T_{\uparrow },
\end{eqnarray*}
%
%
where $R_{\uparrow }$ and $R_{\downarrow }$ are the reflection
coefficients with an incident spin up or down. For the unpolarized
statistical mixture the flux conservation implies that at the
resonances
%
%
\[
T=\frac{1}{2}\left( T_{\uparrow }+T_{\downarrow }\right) =R=\frac{1}{2}\left(
R_{\uparrow }+R_{\downarrow }\right) =\frac{1}{2}. 
\]
%
%
%
%
\begin{figure}
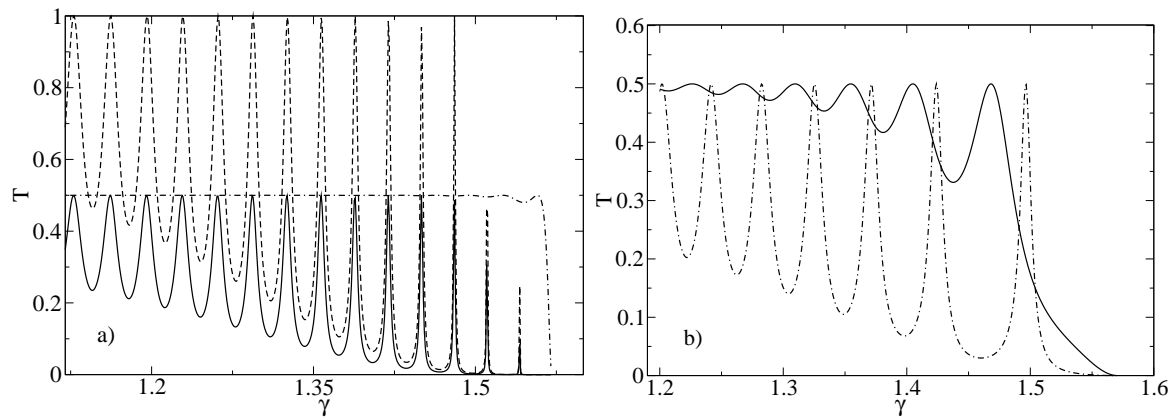

	\centering \includegraphics[width=3.in]{figure3a.eps}
	\includegraphics[width=3.in]{figure3b.eps} \caption{Panel a):
	The transmission coefficient for unpolarized electrons as
	function of the injection angle. The solid line is for
	$u_1/k^2=0.1$, $u/k=0.4$ and $k_0/k=0.05$, the dashed line is for
	$u_1/k^2=0$, $u/k=0.4$ and $k_0/k=0$ and finally the
	dotted-dashed line is for $u_1/k^2=0.1$, $u/k=0.4$ and
	$k_0/k=0.05$. \\ Panel b): The transmission coefficient for
	unpolarized electrons as function of the injection angle. The
	solid line is for $u_1/k^2=0.05$, $u/k=0.0$ and $k_0/k=0.05$ and the
	dotted-dashed line is for $u_1/k^2=0.05$, $u/k=0.4$ and
	$k_0/k=0.05$. \label{figure7}}
\end{figure}
%
%

The calculated transmission coefficients for unpolarized ($T\left( \gamma
\right)$) and polarized ($T_{\downarrow }\left( \gamma \right)
,T_{\uparrow }\left( \gamma \right)$) injected electrons are shown
in Fig.~(\ref{figure2}) with realistic parameters
\cite{Nitta,schapers,grundler}.  For the inverted
In$_{0.53}$Ga$_{0.47}$As/In$_{0.52}$Al$_{0,48}$As heterostructure the
value of the Fermi wave vector is $k=3.53\times 10^6$cm$ ^{-1}$, the
effective mass is $m=0.05$ $m_e,$ whereas the strength of the SO $
k_0$ ranges from $0.01\,k$ to $0.05\,k$. The main feature of
$T_{\downarrow },T,T_{\uparrow }$ is the double step that originates
from the spin-double refraction. When $\gamma $ goes over $\gamma _0$
only the wave (--) can reach the second interface and the transmission
coefficients tend to halve themself. The panel a) of the
Fig.~(\ref{figure2}) shows that, when $\gamma >\gamma _0$, the
resonances of the unpolarized transmission $T$ have height $1/2$. At
the offset $u_1/k^2=2k_0/k$ a limit angle $\gamma_1$ appears also for
the the wave (--). Increasing the offset at higher values both the
limit angles $\gamma_0$ and $\gamma _1$ tend to zero.  At the offset
$2k_0k$ the second step of the transmission becomes almost perfectly
squared as the panel b) of the Fig.~(\ref{figure2}) shows.  At this
optimum value, we have $k_{-}\equiv k$ and $\beta\equiv\gamma$: the
mode (--) is no more refracted. With $u=0$ and $\gamma$ greater than
$\gamma_0$ the resonances within the cavity disappear.  

Since the output spin polarization  of the electrons entering with
spin up or down coincides at any angle $\gamma $ with $T_{\uparrow }$
or $T_{\downarrow}$ then the crossing of R zone depolarizes the
electrons. The polarization $P_{\text{out}}$ of the electrons entering
unpolarized is equal to $T$ only for $
\gamma >\gamma_0$, and for $\gamma <\gamma _0$ it tends to zero for $\gamma
\rightarrow 0$: the crossing of R zone gives rise to a spin polarization
that is absent at normal incidence. At the optimum value of the offset
$u_{1} = 2 k k_{0}$ and for $\gamma>\gamma_0$, $P_\text{out}$ is
independent of the incidence angle. We note that this feature is
robust with respect to few elastic scattering events that conserve $k$
but change its direction. This behavior reminds the non-ballistic
spin-FET proposed by Schliemann {\it et al.}~\cite{schliemann}.

Is is interesting to do an analysis on the proper modes of the Rashba
region. In Fig.~(\ref{figure7}) is shown the transmission as function
of the injection angle for $\gamma>\gamma_0$ and several values of the
parameters. In the Panel a) of Fig.~(\ref{figure7}) the dashed curve
corresponds to the proper mode of the cavity without SO coupling
($u_1/k^2=0$, $u/k=0.4$ and $k_0/k=0$). If we now turn on the SO
coupling and choose the offset to the optimum value $u_1= 2k_0k$
(solid curve -- $u_1/k^2=0$, $u/k=0.4$ and $k_0/k=0.05$) we observe
the halving of the transmission maxima but those preserve the proper
modes of the cavity. The proper modes of the cavity are washed out for
a perfectly transparent the cavity at the optimum offset (dotted-dashed
curve -- $u_1/k^2=0.1$, $u/k=0$ and $k_0/k=0.05$). In the last case
the flatness of the transmission coefficient (and of polarization
vector) is independent by effects related to multiple reflections
inside the cavity. In the Panel b) of Fig.~(\ref{figure7}) are
reported similar results with the offset to a value lower than the
optimum one demonstrating how a perfectly transparent barrier ($u=0$)
has proper modes due only to SO coupling and how this modes modifies
when $u$ increases.

%
%
\begin{figure}
	\centering \includegraphics[width=3.in]{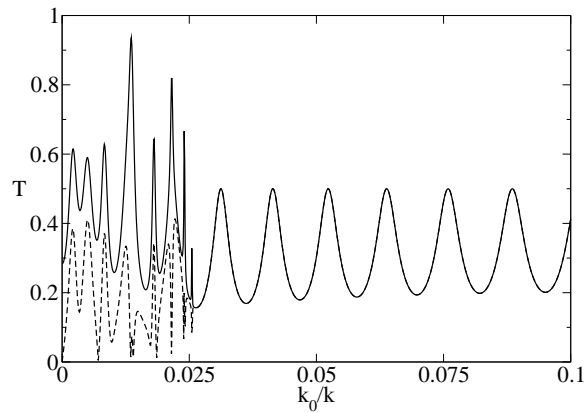} \caption{
	The transmission coefficient for unpolarized electrons when
	$\gamma =1.25$. The threshold $\overline{k_0}$ is
	$0.025k$. The dotted curve gives the output spin polarization,
	$L=283$ nanometers and we have put $u/k=0.4$ to enhance the
	resonances in the low step of the transmission. \label{figure3}}
\end{figure}
%
%

We stress that the dependence of the limit angle $\gamma _0$ on the SO
strength $k_0$ suggests the possibility to build up a spin-FET
operating on spin unpolarized electrons injected in the R region. The
electrons emerge in the NR drain region partially polarized with a
polarization controlled by a gate electrode via the SO
interaction. There is a SO strength $\overline{k_0} $ at which $\gamma
=\gamma _0$. The Fig.~(\ref{figure3}) shows that the transmission
coefficient exhibits irregular Fabry-Perot oscillations below $
\overline{k_0}$, whereas for $k_0>\overline{k_0}$ the oscillations become
regular with maxima equal to $1/2.$ The threshold $\overline{k_0}$ is
determined by the offset 
%
%
\[
u_1/k^2=2\overline{k_0}/k, 
\]
%
%
for which $\gamma _0$ goes over $\gamma $ and the wave (+) propagation
ceases. When $k_0>\overline{k_0}$ the polarization of electrons in the
NR drain $P_{\text{out}}$ is equal to $T$ as we have seen before. We
get a source of a spin polarized current controlling $k_0$ with a
gate.

We notice that Mireles and Kirczenow \cite{Mireles} have studied the
scattering against a finite Rashba region within a quantum wire.  They
show that, injecting spin up electron, the ballistic spin conductance
oscillates varying the SO coupling strength and they claim that these
results may be of relevance for the implementation of quasi-one
dimensional spin transistor device. We have shown that a ballistic
conductance oscillation with $k_0$ appear also without lateral
confinement and without ferromagnetic source and drain, that is by
handling spin unpolarized electrons.

Another issue that deserves some attention is the inclusion of the
linear term due to Dresselhaus spin-orbit
coupling~\cite{dresselhaus}. There is a recent experimental
evidence~\cite{ganichev} that the Dresselhaus coupling in some
heterostructures can be of the same order of the Rashba coupling, up
to one half of Rashba strength. Some preliminary results for a single
interface with a Rashba zone indicate that, even in the less
favourable case of normal incidence, the addition of Dresselhaus term
everywhere in the 2DEG plane gives different transmissions for the two
orthogonal incident spin states. In our opinion the further inclusion
of the Dresselhaus spin-orbit coupling should even enhance the
polarization effects that we have found with the only Rashba term.

\section{Phase averaging due to the thermal broadening}

%
%
\begin{figure}
	\centering \includegraphics[width=6.in]{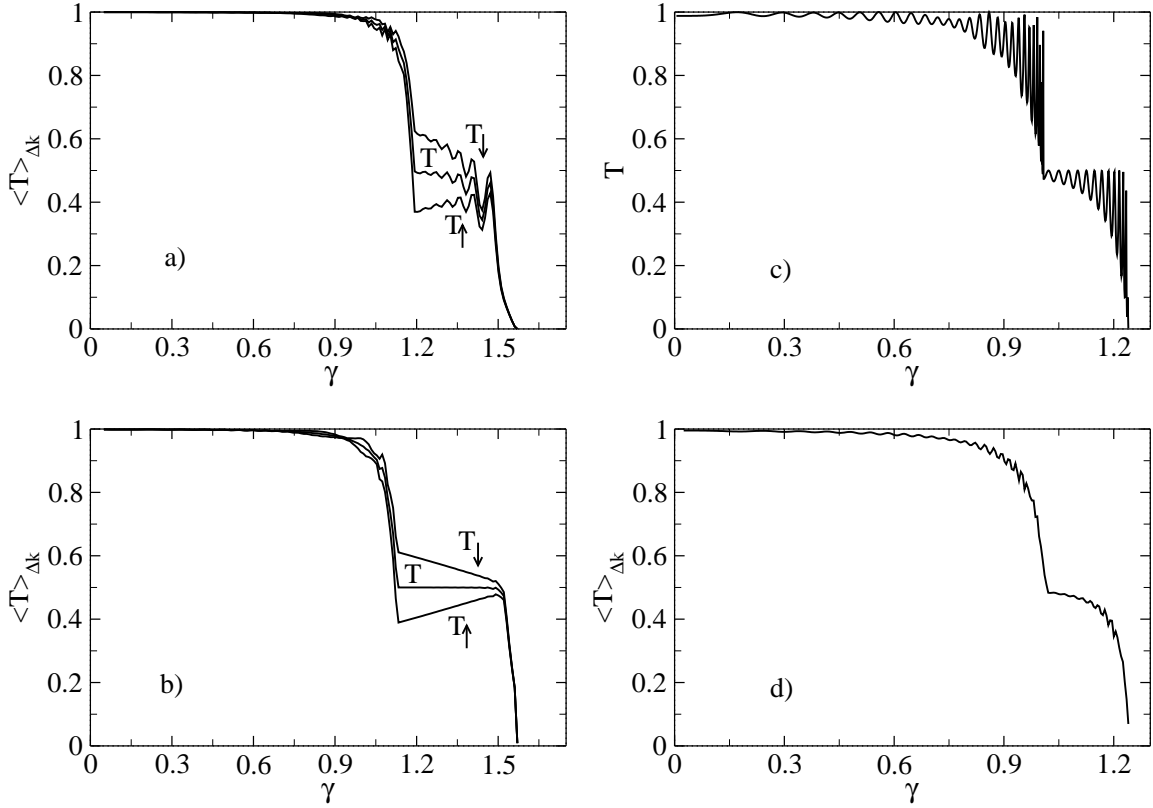}
	\caption{Panel a): Thermal-averaged transmission coefficient
	for unpolarized electrons as function of the injection
	angle. The curves are relatives to $u_1/k^2=0.05$, $u/k=0$ and
	$k_0/k=0.05$. The thermal average corresponds to a temperature
	of 5 K.  \\ Panel b): As in the Panel a) but with
	$u_1/k^2=0.1$, $u/k=0$ and $k_0/k=0.05$. The thermal average
	corresponds to a temperature of 5 K. \\ Panel c): Transmission
	coefficient at 0 K for unpolarized electrons as function of the
	injection angle with $u_1/k^2=0.2$, $u/k=0$ and
	$k_0/k=0.05$. \\ Panel d): Thermal-averaged transmission
	coefficient for unpolarized electrons as function of the
	injection angle. The curve is relative to $u_1/k^2=0.2$,
	$u/k=0$ and $k_0/k=0.05$. The thermal average corresponds to a
	temperature of 5 K.  \label{figure4}}
\end{figure}
%
%
%
%
\begin{figure}
	\centering \includegraphics[width=3.in]{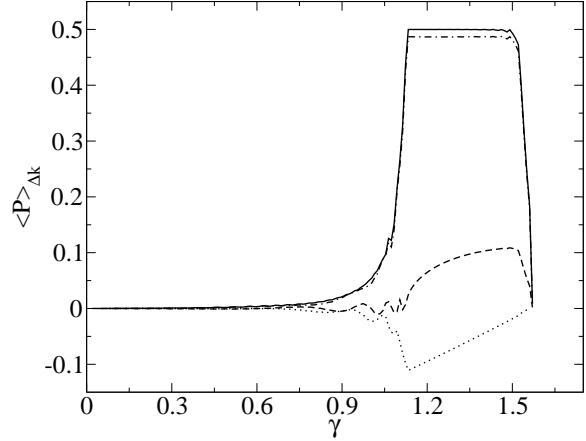}
	\caption{Thermal-averaged polarization as function of the
	injection angle. The solid curve is the module of the
	polarization vector, the dashed-dotted line corresponds to
	$<\sigma_x>$, the dashed line corresponds to $<\sigma_y>$ and
	the dotted line corresponds to $<\sigma_z>$. The curves are
	evaluated for $u_1/k^2=0.1$, $u/k=0$ and $k_0/k=0.05$. The
	thermal average corresponds to a temperature of 5 K.
	\label{figure5}}
\end{figure}
%
%

Now we want to take into account the smoothing effect of the Fermi
surface due to finite temperatures. Supposing that we are in the
linear response regime we have for the current $I$
%
%
\[
I=G\left( E_\text{F}\right) \frac{\mu _1-\mu _2}e
\]
%
%
$\mu _1-\mu _2$ being the applied bias and $G\left(
E_\text{F}\right) $ the conductance ($\mu _1-\mu
_2<<E_\text{F}$). For a ballistic conductor \cite{Datta}
%
%
\[
G\left( E_\text{F}\right) =\frac{2e^2}h\int T\left( E\right)
F_\text{T}\left( E-E_\text{F}\right) dE
\]
%
%
where $F_\text{T}\left( E-E_\text{F}\right) $ is the thermal
broadening function
%
%
\[
F_\text{T}\left( E-E_\text{F}\right) =-\frac d{dE}\frac 1{\exp [\left(
E-E_\text{F}\right) /k_\text{B} 
\overline{T}]+1}=\frac
1{4k_\text{B}\overline{T}}\textrm{sech}^2\left( \frac
E{2k_\text{B}\overline{T}}\right)
\]
%
%
$\overline{T}$ being the temperature. The thermal average of the
transmission is
%
%
\[
\left\langle T\right\rangle _{\text{th}}=\int T\left( E\right)
F_\text{T}\left( E-E_\text{F}\right) dE.
\]
%
%
If we approximate the Fermi-Dirac distribution with the ramp 
%
%
\[
\left\{ 
\begin{array}{c l}
1 & E<E_\text{F}-2k_\text{B}\overline{T} \\ 
\frac 12-\left( E-E_\text{F}\right) /4k_\text{B}\overline{T} &
E_\text{F}-2k_\text{B}\overline{T}
<E<E_\text{F}+2k_\text{B}\overline{T} \\ 
0 & E>E_\text{F}+2k_\text{B}\overline{T}
\end{array}
\right. 
\]
%
%
we get 
%
%
\begin{equation}
\left\langle T\right\rangle _{\text{th}}=\frac 1{2\Delta
k}\int_{k_\text{F}-\Delta 
k}^{k_\text{F}+\Delta k}T\left( k\right) dk \label{10}
\end{equation}
%
%
where we assume that 
%
%
\[
\Delta
k=\frac{k_\text{B}\overline{T}}{E_\text{F}}k_\text{F}<<k_\text{F}.
\]
%
%
With a Fermi energy $E_\text{F}=14$ meV, and $\overline{T}=3$ K we
have $\Delta k/k_\text{F}=0.018$ and in the following we choose
$\Delta k/k_\text{F}=0.03$ corresponding to a temperature of 5 K. We
start by considering the thermal average of the transmissions. In the
Fig.~(\ref{figure4}) we show the averaged transmission for various
value of the offset. In the Panel a) of Fig.~(\ref{figure4}) the
offset is $u_1/k^2=0.05$ is  and few resonances appear when $\gamma$
is larger than $\gamma_0$; increasing the temperature up to 5 K the
rapid oscillations of the transmission below $\gamma_0$ are almost
completely canceled whereas the resonances of the (--) mode are still
well defined in the averaged transmission. In the Panel b) of
Fig.~(\ref{figure4}) the offset is chosen at the optimum value
$u_1/k^2=2k_0/k=0.1$. We observe again an almost perfectly squared
transmission step among $\gamma_{0}$ and $\pi/2$ , whereas below
$\gamma_{0}$ the rapid Fabry-Perot oscillations are washed out. The
transmission steps for a larger offset $u_1/k^2=0.2$ at 0 K and at 5 K
are compared in the Panel c) and d) of Fig.~(\ref{figure4}). In this
case the transmission oscillations are canceled at any incidence
angle but the double step structure survives. 

The thermal averaging has been performed also on the polarization
vector and the Fig.~(\ref{figure5}) shows the average calculated with
Eq.~(\ref{10}) of the modulus of the polarization vector and of its
three components. The offset is at the optimum value and the output
spin-polarization tends to be nearly orthogonal to the interfaces as
one expect because $\vec{k}_{-}$ has a small component in $x$
direction and the spin and the momentum are orthogonal each other. 

\begin{figure}
	\centering \includegraphics[width=3.in]{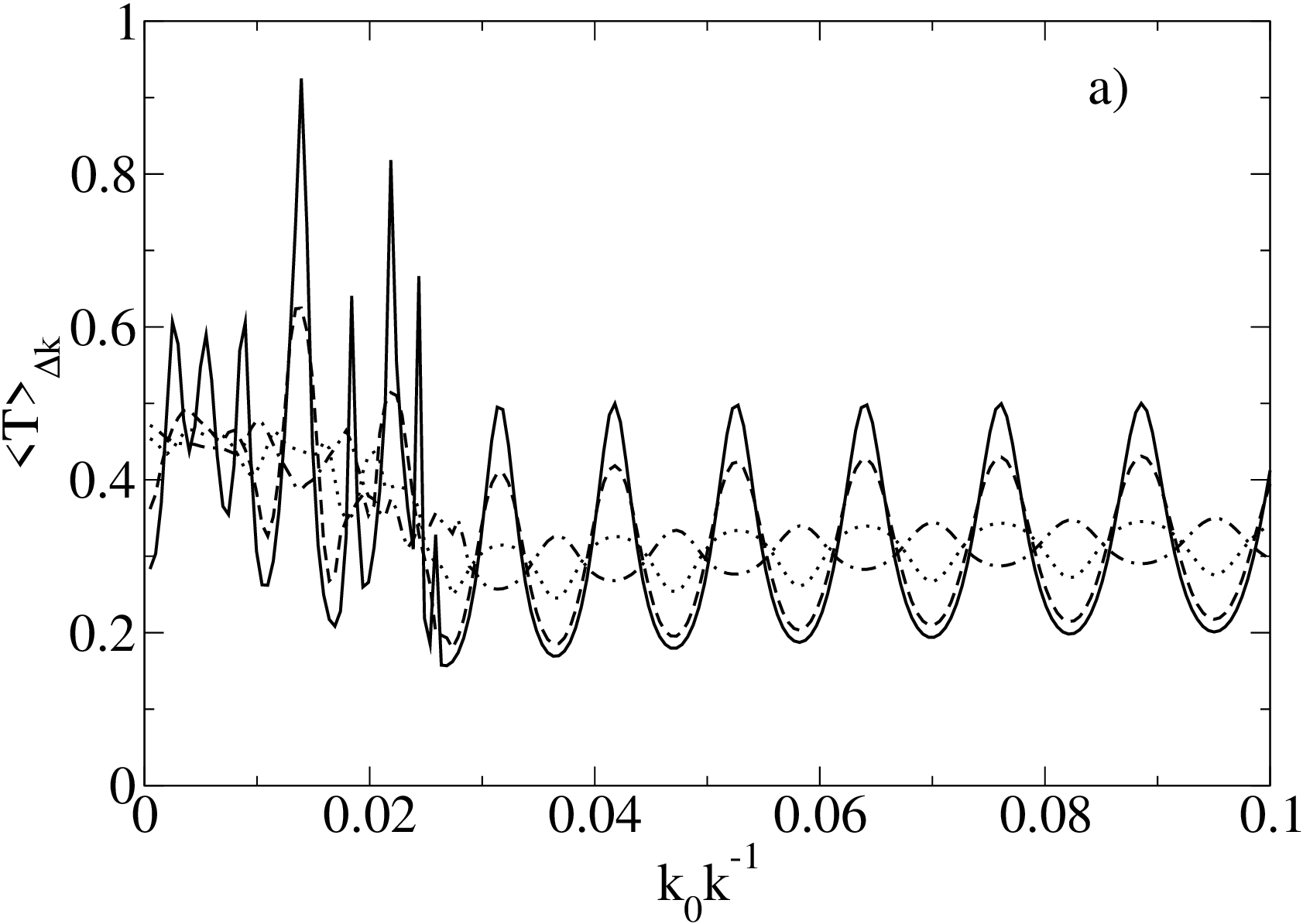}
	\includegraphics[width=3.in]{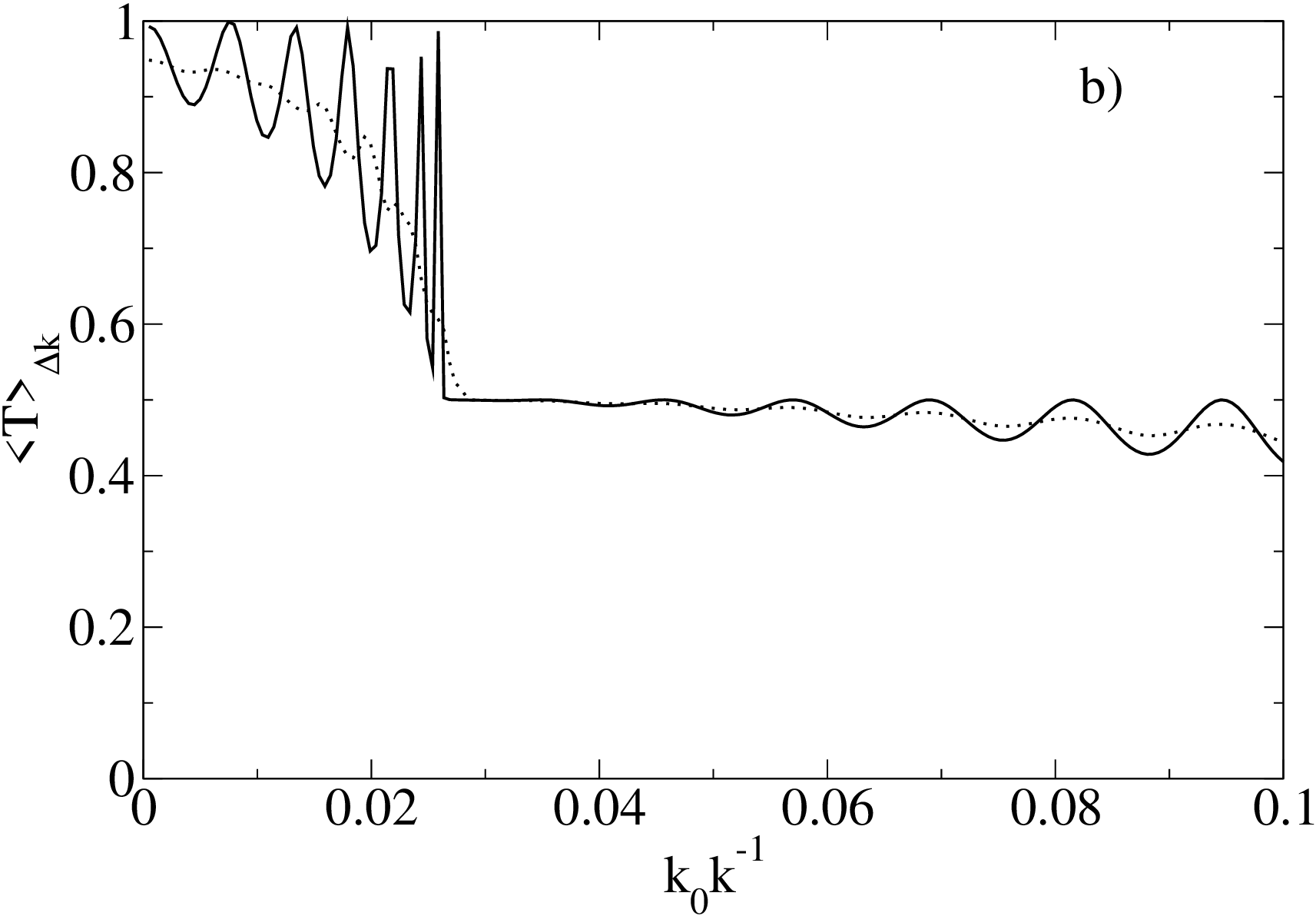} \caption{The
	thermal-averaged transmission coefficient for unpolarized
	electrons when $\gamma =1.25$ as function of the spin-orbit
	coupling. The threshold $\overline{k_0}$ is $0.025k$. \\ Panel
	a): The solid curve corresponds to zero temperature, the
	dashed line to a temperature of 3 K, the dotted line to a
	temperature of 5 K and the dashed-dotted to a temperature of 8
	K. \\ Panel b): The thermal-averaged transmission coefficient
	for unpolarized electrons when $\gamma =1.25$ as function of
	the spin-orbit coupling at 5 K (solid line) and 8 K (dotted
	line), with perfectly transparent interfaces ($u=0$).
	\label{figure6}}
\end{figure}

The Fig.~(\ref{figure6}) shows the thermal average of the transmission
of unpolarized electrons as a function of the SO strength with the
same parameter of the Fig.~(\ref{figure3}) where the transmission has
been calculated at 0 K. Therefore we judge that an increase of few
Kelvin degrees does not cancel the modulation effect.
%
%
%
%

\section{Conclusions}

We have shown that the spin-double refraction in NR--R--NR system
allows both the current modulation and the polarization of
spin-unpolarized injected electrons. Furthermore the existence of
limit angles for the propagation within the structure involves the
halving of the transmission and this could searched for a sign of the
double refraction phenomenon. About the feasibility of the studied
hybrid system we stress that our calculations use realistic parameter
values for the Rashba SO coupling~\cite{Nitta}. We think that the
physics of spin-double refraction and its use to produce and control
spin-polarized currents should be within the present experimental
investigations. We propose that the injection of an electron beam with
an incidence angle on a Rashba barrier greater than the limit angle
gives an output current modulated by the SO strength. Our proposal is
different from the original one of Datta and Das~\cite{Datta-Das} in
which the current modulation stems out from the interference of both
the modes (+) and (--), that is from the precession within the Rashba
region. Instead we propose to use the spin rotation that appears when
an electron beam goes through the interfaces with a incident angle
$\gamma$ out of the normal.  Such angle could be realized by using an
adiabatic quantum point contact as source in which the constriction
axis forms the angle $\gamma$ with the NR-R interface.

\ack
We gratefully acknowledge helpful discussions with I. D'Amico,
D. Frustaglia and M. Governale.

\appendix
\section{Output spin polarization evaluation}\label{appendix:I}

In this appendix we give the proof of the relation (\ref{9}). In the
general case, when we work with density matrix operator 
%
%
\begin{equation*}
\rho = \sum_m | m \rangle p_m \langle m |, 
\end{equation*}
%
%
the average value of the observable $A$ is the trace of $\rho A$:
%
%
\begin{equation*}
\langle A \rangle = \Tr \left[ \rho A \right],
\end{equation*}
%
%
that in terms of the density matrix elements is equal to
%
%
\begin{equation*}
\Tr\left[ \rho A \right] = \sum_{m} p_{m} ( | m \rangle
\langle m | A ) = \sum_m p_m \langle m | A | m \rangle.
\end{equation*}
%
%
In present case the density matrix operator has been defined in the
Eq.~\ref{dario_one} with the components (\ref{dario_ol}). The averaged
value of the modulus of the polarization is defined by
%
%
\begin{equation}\label{a:one}
\left| \vec{P}_{\text{out}} \right| = \sqrt{\langle \sigma_x \rangle^2 +\langle
\sigma_y \rangle^2 +\langle \sigma_z \rangle^2}.
\end{equation}
%
%
Using the density matrix operator (\ref{dario_one}) we have that
%
%
\begin{equation}\label{a:two}
\begin{array}{r c l}
\langle \sigma_x \rangle & = & \displaystyle\frac{1}{2}
\left(t_{\uparrow\uparrow}^* 
t_{\downarrow\uparrow} + t_{\downarrow\uparrow}^*
t_{\uparrow\uparrow} + t_{\uparrow\downarrow}^*
t_{\downarrow\downarrow} + t_{\downarrow\downarrow}^*
t_{\uparrow\downarrow}\right) \\
\langle \sigma_y \rangle & = & -\displaystyle\frac{i}{2} \left(
t_{\uparrow\uparrow}^* 
t_{\downarrow\uparrow} - t_{\downarrow\uparrow}^*
t_{\uparrow\uparrow} + t_{\uparrow\downarrow}^*
t_{\downarrow\downarrow} - t_{\downarrow\downarrow}^*
t_{\uparrow\downarrow} \right)\\
\langle \sigma_z \rangle & = & \displaystyle\frac{1}{2} \left( \left| 
t_{\uparrow\uparrow} \right|^2 
-\left| t_{\downarrow\uparrow} \right|^2 + \left|
t_{\uparrow\downarrow} \right|^2 - \left| t_{\downarrow\downarrow}
\right|^2 \right) .
\end{array}
\end{equation}
%
%
Substituting the relations (\ref{a:two}) into (\ref{a:one}) we have
that: 
%
%
\begin{equation}\label{a:three}
\begin{array}{r c l}
\left| \vec{P}_\text{out} \right|^2 & = & \frac{1}{4}\left[ \left|
t_{\uparrow\uparrow} 
\right|^4 +  \left| t_{\downarrow\uparrow} \right|^4 + 
\left| t_{\uparrow\downarrow} \right|^4 + \left| t_{\downarrow\downarrow}
\right|^4 + 2\left( \left| t_{\uparrow\uparrow}\right|^2 \left|
t_{\downarrow\uparrow}\right|^2 + \left| t_{\uparrow\uparrow}\right|^2 \left|
t_{\uparrow\downarrow}\right|^2  \right. \right.\\
&& - \left| t_{\uparrow\downarrow}\right|^2 \left|
t_{\downarrow\uparrow}\right|^2 + \left| t_{\downarrow\downarrow}\right|^2 \left|
t_{\downarrow\uparrow}\right|^2 + \left| t_{\downarrow\downarrow}\right|^2 \left|
t_{\uparrow\downarrow}\right|^2 - \left| t_{\downarrow\downarrow}\right|^2 \left|
t_{\uparrow\uparrow}\right|^2 \\ && \left.\left.
+2 t_{\downarrow\downarrow} t_{\uparrow\uparrow}
t_{\downarrow\uparrow}^* t_{\uparrow\downarrow}^* + 2
t_{\downarrow\downarrow}^* t_{\uparrow\uparrow}^* 
t_{\downarrow\uparrow} t_{\uparrow\downarrow} \right)\right]
\end{array}
\end{equation}
%
%
When the injection angle $\gamma$ approaches to zero, we have that the
system does not flip the spin, that is $t_{\uparrow\downarrow} =
t_{\downarrow\uparrow} = 0$ so that 
%
%
\begin{equation}\label{app:enzo}
\left| \vec{P}_\text{out} \right| = \frac{1}{2} \left|  \left|
t_{\uparrow\uparrow} \right|^2 - \left|
t_{\downarrow\downarrow} \right|^2  \right|,
\end{equation}
%
%
that is in the case of injection of unpolarized electrons
the output polarization is zero.

We note that in the R zone the spin state of the $(+)$ and $(-)$ modes
given by the spinors $|1\rangle$ and $|2\rangle$ are
conserved~\cite{Marigliano}. The output spin state is strictly
determined by the transmitted amplitudes of the interfaces. The
interference between the mode $(+)$ and $(-)$ when both are present
($\gamma<\gamma_0$), makes the polarization different from the
transmission as the Eq. (\ref{app:enzo}) shows.
When $\gamma>\gamma_0$ only the $(-)$ mode survives in the R zone.
This is demonstrated by the fact that the inner product between
the state $|1\rangle$ and $|2\rangle$ goes to one
%
%
\begin{equation*}
\left| \langle 1 | 2 \rangle \right| = 1 \quad \textrm{for} ~ \gamma>\gamma_0,
\end{equation*}
%
%
this means that the two wave functions, for $\gamma>\gamma_0$, differ
by a phase factor
%
%
\begin{equation}
\left| 1\right\rangle = e^{-i\phi} \left| 2 \right\rangle \quad\to\quad \frac
{1}{\sqrt{T_{\uparrow }}} 
{t_{\uparrow \uparrow } \choose t_{\downarrow \uparrow}} =
 e^{-i\phi}  \frac{1}{\sqrt{
T_{\downarrow }}}
{t_{\uparrow \downarrow} \choose t_{\downarrow \downarrow}}.
\end{equation}
%
%
Using the previous relation we can express $t_{\uparrow\downarrow}$
and $t_{\downarrow\uparrow}$ as function of $t_{\uparrow\uparrow}$ and
$t_{\downarrow\downarrow}$:
%
%
\begin{equation}\label{a:four}
t_{\downarrow\uparrow} = e^{i\phi}\sqrt{
\displaystyle\frac{T_\downarrow}{T_\uparrow}} t_{\uparrow\uparrow}
\quad \textrm{and} \quad t_{\uparrow\downarrow} = e^{-i\phi}\sqrt{
\displaystyle\frac{T_\uparrow}{T_\downarrow}} t_{\downarrow\downarrow}.
\end{equation}
%
%
Substituting those expressions in the (\ref{a:three}) we have 
%
%
\begin{equation}
\left| \vec{P}_\text{out} \right| = \frac{1}{2} \left[\left|
t_{\uparrow\uparrow} 
\right|^2 + \left| t_{\downarrow\downarrow} \right|^2 +
\displaystyle \frac{T_\uparrow^{\,2}}{T_\downarrow^{\,2}} 
\left| t_{\downarrow\downarrow} \right|^2 +
\displaystyle\frac{T_\downarrow^{\,2}}{T_\uparrow^{\,2}} 
\left|t_{\uparrow\uparrow} \right|^2 \right]
\end{equation}
%
%
and using the relations (\ref{a:four}) this is equivalent to
%
%
\begin{equation}
\left| \vec{P}_\text{out} \right| = \frac{1}{2} \left[ \left|
t_{\uparrow\uparrow} 
\right|^2 + \left| t_{\downarrow\downarrow} \right|^2 + 
\left| t_{\uparrow\downarrow} \right|^2 +
\left|t_{\downarrow\uparrow} \right|^2 \right] = \frac{T_\uparrow +
T_\downarrow}{2} = T.
\end{equation}
%
%


\Bibliography{99}

\bibitem{awschalom}  {\it Semiconductor Spintronics and Quantum
Computation} , edited by D.D. Awschalom, D. Loss and N. Samarth
(Spinger, Berlin 2002).

\bibitem{zutic} I. \v Zut\'\i c, J. Fabian, and S. Das Sarma,
Rev. Mod. Phys. {\bf 76}, 323 (2004).

\bibitem{Datta-Das}  S. Datta, and B. Das, Appl. Phys. Lett. {\bf 56}, 665
(1990).

\bibitem{Rashba}  E.I. Rashba, Fiz. Tverd. Tela (Leningrad) {\bf 2},1234 (1960)
[Sov. Phys. Solid State {\bf 2},1109 (1960)].

\bibitem{Schmidt}  G. Schmidt, D. Ferrand, L.W. Molenkamp, A.T. Filip and
B.J. Van Wees, Phys. Rev. B {\bf 62}, 4790(R) (2000).

\bibitem{Kiselev}  A.A. Kiselev, and K.W. Kim, Appl. Phys. Lett {\bf 78}, 775
(2001).

\bibitem{Koga}  T. Koga, J. Nitta, H. Takayanagi and S. Datta,
Phys. Rev. Lett.{\bf 88}, 126601 (2002).

\bibitem{Egnes}  J.C. Egnes, G. Burkard, and D.Loss, Phys. Rev. Lett. {\bf 89},
176401 (2002).

\bibitem{jiang} Y. Jiang, and M.B. Jalil, J. Phys.: Condens. Matter {\bf
15}, L31 (2003).

\bibitem{frustaglia} D. Frustaglia, and K. Richter, Phys. Rev. B {\bf
69}, 235310 (2004).  

\bibitem{Pala}  M.G. Pala, M. Governale, J. K\"{o}nig, U. Z\"{u}licke,
Europhys. Lett. {\bf 65}, 850 (2004). 

\bibitem{Ion}  Radu Ionicioiu and Irene D'Amico, Phys. Rev.B {\bf 67, }
041307(R) (2003).

\bibitem{Govpump}  M. Governale, F. Taddei and Rosario Fazio,
Phys. Rev. B {\bf 68}, 155324 (2003).

\bibitem{schliemann}  J. Schliemann, J. Carlos Egues and D. Loss, Phys. Rev.
Lett. {\bf 90}, 146801 (2003).

\bibitem{dresselhaus}  G. Dresselhaus, Phys. Rev. {\bf 100}, 580 (1955).

\bibitem{Marigliano}  V. Marigliano Ramaglia, D. Bercioux, V. Cataudella, G.
De Filippis, C.A. Perroni and F. Ventriglia, Eur. Phys. J. B {\bf 36}, 365
(2003).

\bibitem{Khodas} M. Khodas, A. Shekhter, A.M. Finkel'stein,
Phys. Rev. Lett. {\bf 92}, 086602,(2004).

\bibitem{Nitta}  J. Nitta, T. Akazaki , H. Takayanagi and T. Enaki,
Phys. Rev. Lett. {\bf \ 78}, 1138 (1997).

\bibitem{schapers}  T. Sch\"{a}pers, J. Engels, T. Klocke, M. Hollfelder
and H. L\"{u}th, J. Appl. Phys. {\bf 83}, 4324 (1998).

\bibitem{grundler}  D. Grundler, Phys. Rev. Lett. {\bf 84}, 6074 (2000).

\bibitem{urRashba}  Y.A. Bychkov and E.I. Rashba, J. Phys. C {\bf 17}, 6039
(1984).

\bibitem{Ferry}  D.K. Ferry, and S.M. Goodnick, {\sl Transport in
Nanostructures} (Cambridge University Press, Cambridge 1997).

\bibitem{note1} The first arrow in the label of
the transmitted amplitudes represents the output spin, whereas the
second arrow indicates the input spin.

\bibitem{Mireles}  F. Mireles, and G. Kirczenow, Phys. Rev. B {\bf 
64}, 024426 (2001).

\bibitem{ganichev} S. D. Ganichev, V. V. Bel'kov, L. E. Golub,
E. L. Ivchenko, P. Schneider, S. Giglberger, J. Eroms, J. De Boeck,
G. Borghs, W. Wegscheider, D. Weiss, and W. Prettl 
Phys. Rev. Lett. \textbf{92}, 256601 (2004).

\bibitem{Datta}  S. Datta, {\sl Electronic Transport in Mesoscopic
Systems }(Cambridge University Press, Cambridge 1995).
\end{thebibliography}
\end{document}